\documentstyle[12pt]{article}

\newcommand{\eq}[1]{eq.(\ref{#1})}
\newcommand{\dpar}[2]{\frac{\partial #1}{\partial #2}}

\def\be{\begin{equation}}
\def\ee{\end{equation}}
\def\a{\alpha}

\def\log{\mbox{ln}}
\def\Tr{\mbox{Tr}}

\begin{document}
\begin{flushright}
UCLA/93/TEP/41 \\
November 1993 \\
hep-ph/9311229
\end{flushright}
\vspace{0.5in}
\begin{center}
{\Large Effective action of magnetic monopole in
three-dimensional electrodynamics with massless matter
and gauge theories of superconductivity} \\
\vspace{0.4in}
{\large S.Yu.~Khlebnikov}
\footnote{On leave of absence
from Institute for Nuclear Research of the Academy
of Sciences, Moscow 117312 Russia.
Address after 1 December 1993: Department of Physics, Purdue University,
West Lafayette, IN 47907. } \\
\vspace{0.2in}
{\it Department of Physics, University of California,
Los Angeles, CA 90024, USA} \\
\vspace{0.7in}
{\bf Abstract} \\
\end{center}
We compute one-loop effective action of magnetic monopole in
three-dimensional electrodynamics of massless bosons and fermions and
find that it contains an infrared logarithm.
So, when the number of massless matter species is sufficiently large,
monopoles are suppressed and in the weak coupling limit charged particles
are unconfined. This result provides some support to gauge theories
of high-temperature superconductors.
It also provides a mechanism by which interlayer
tunneling of excitations with one unit of the ordinary electric charge
can be suppressed while that of a doubly charged object is allowed.
\newpage
Gauge theories of high-temperature superconductors \cite{BA}
assume that in certain planar electronic systems spin and charge are
separated and the resulting new quasiparticles interact via abelian
gauge forces. This idea is seemingly in contradiction with
three-dimensional confinement due to magnetic monopoles \cite{Polyakov}.
One may try to resolve the contradiction by
assuming that some of the charged excitations are gapless. Then
it is possible that quantum effects of these excitations are sufficiently
strong in the infrared to modify the interaction between monopoles in such
a way that confinement of charges is lost. To see how this can happen,
consider the relativistic version of the problem --
three-dimensional quantum electrodynamics of massless bosons or
fermions. A simple calculation shows that one-loop contribution of
massless charged particles to the gauge field propagator causes
its small momentum behavior to change from the usual $1/p^2$ to $1/|p|$.
The bilinear part of the corresponding effective action for the gauge field
calculated on the monopole configuration then produces a logarithm of
the total size of the system. Indeed, the field strength of monopole is
$F_{ij}\sim \epsilon_{ijk} x_k / r^3$, therefore,
\be
\int d^3 x F_{ij} \frac{1}{\sqrt{\partial^2}} F_{ij}
\sim \log\frac{R}{a} \; ,
\label{bil}
\ee
where $R$ and $a$ are infrared and ultraviolet cutoffs, respectively.
Essentially the same calculation appears in ref.\cite{IL}.
The coefficient of the logarithm increases proportionally to the number
$N$ of massless species.
This suggests that in the presence of massless charged particles,
at least when $N$ is large enough,
monopoles are suppressed via a version of the "infrared
catastrophe" and charged particles are unconfined.

The reason why \eq{bil} is not sufficient to
determine the fate of magnetic monopoles in the presence of massless
charged particles
even in the large $N$ limit, is that the interaction of
a charged particle with monopole has no small parameter, so
the bilinear part (\ref{bil}) of the effective action is in no way
distinguished relative to terms containing more
powers of the gauge field. To make a reliable conclusion, we need all
these terms. On the other hand, because the gauge propagator in the
large $N$ limit is of order $1/N$, the large $N$ limit suppresses
higher-loop contributions to the effective action. Therefore,
in this limit the problem reduces to calculation of the one-loop
determinants of massless bosons and fermions in the monopole background.
For this calculation, we chose to proceed with relativistic three-dimensional
particles. Precise dispersion laws for quasiparticles that may occur in real
electronic systems are unknown at present.
The result and the main steps of our calculation are
presented below. The result confirms the presence of an infrared
logarithm in monopole's effective action both in bosonic and
fermionic cases. We thus show that single monopoles
are suppressed by an "infrared catastrophe" in the presence of a
sufficient number of massless matter fields. We cannot state
at present what exactly this "sufficient" number is because the answer
to this question lies outside the region of validity of the large
$N$ approximation.

Consider now the weak coupling limit of
three-dimensional electrodynamics when the dimensionful gauge coupling
$e^2$ is much smaller
than the ultraviolet scale $M\sim a^{-1}$ at which internal structure of
monopole becomes essential. Then, in the absence of massless matter, monopoles
and anti-monopoles would form a dilute gas. When the number of
massless matter species is sufficiently large, the infrared logarithm
causes monopoles and anti-monopoles to assemble into
"molecules" -- pairs of typical size $d$ that is much smaller than the
average distance between the pairs. These pairs interact by a short range
potential of order $(d^2/r^2) \log (d/r)$, so it is natural to expect
that charged particles are unconfined.
This picture provides some support to the idea of new gauge interactions
in planar electronic systems.
At the end of this paper we discuss some further applications of our results.

The one-loop contributions to monopole's effective action from
a single charged bosonic field and a single charged fermionic field
are respectively of the form
\be
S^{(1)}_B = \Tr\log (-D^2) - \Tr\log(-\partial^2) \; ,
 ~~~~~~~S^{(1)}_F =
- \Tr\log (\sigma D) + \Tr\log(\sigma\partial) \; ,
\label{S1}
\ee
where $D$ are covariant derivatives and $\sigma$ are Pauli matrices.
Eq.(\ref{S1}) requires both ultraviolet and infrared regularizations.
We compute not expressions (\ref{S1}) directly but rather
\be
S^{(1)}_B(R) = \Tr\log {\cal M}_B - \Tr\log {\cal M}_{B0} \; ,
 ~~~~~~~S^{(1)}_F (R)=
- \frac{1}{2} \left( \Tr\log (-{\cal M}^2_F) -
\Tr\log (-{\cal M}^2_{F0}) \right) \; ,
\label{S2}
\ee
where
\be
{\cal M}_B = - \frac{1}{4R^4} (r^2 + R^2) D^2 (r^2 + R^2) \; ,
 ~~~~~~~- {\cal M}_F^2 =
- \frac{1}{4R^4} (r^2 + R^2) (\sigma D)^2 (r^2 + R^2) \; ,
\label{cal}
\ee
and ${\cal M}_{B0}$ and $-{\cal M}^2_{F0}$ are obtained analogously
from the operators $-\partial^2$ and $-(\sigma\partial)^2$
of the vacuum sector.
This replacement is similar to the one used by 't Hooft in his
four-dimensional instanton calculation \cite{tHooft}.
If the effective action were not infrared sensitive,
the additional factors of $(r^2+R^2)/R^2$ would cancel between vacuum
and non-vacuum contributions in (\ref{S2}). In the present case, we
intend to show that the effective action {\em is} infrared sensitive.
In this case \eq{S2} provides an infrared regularization of \eq{S1},
$R$ being the regulator radius. In what follows we measure
all distances in units of $R$, hence, we set $R=1$.

The eigenvalue equations for operators (\ref{cal}) are
\be
\left(  D^2 + \frac{4\lambda}{(1+ r^2)^2} \right) \Psi_B = 0 \; ,
\label{bos}
\ee
\be
\left( (\sigma D)^2 + \frac{4\lambda}{(1+ r^2)^2} \right) \Psi_F = 0 \; ,
\label{ferm}
\ee
where $\lambda$ stand generically for the eigenvalues.
In bosonic \eq{bos}, radial and angular variables are separated by
$\Psi_B = \psi(r) Y_{q,l,m}(\theta,\phi)$,
where $Y_{q,l,m}$ are the monopole harmonics of ref.\cite{WY}.
One gets the radial equation
\be
\left[ \left( \dpar{}{r} \right)^2 + \frac{2}{r}\dpar{}{r} -
\frac{\a(\a+1)}{r^2} + \frac{4\lambda}{(1+ r^2)^2} \right]
\psi = 0 \; ,
\label{rad}
\ee
where $\a=[(l+1/2)^2 - q^2]^{1/2} - 1/2$, ~$l=|q|,|q|+1,...$,
and the multiplicity of the eigenvalue $\lambda$ is $(2l+1)$.
Parameter $q$ assumes integer and half-integer values as a consequence
of the Dirac quantization condition. All our results depend only on $|q|$,
so in what follows we take $q\geq 0$.
In fermionic \eq{ferm}, the variables are separated by using either
of the three angular dependences $\xi^{(1)}_{jm}$, $\xi^{(2)}_{jm}$,
$\eta_m$ introduced in ref.\cite{KYG}. We find that in all three cases
the resulting radial equations have the same form as \eq{rad}
but with different values of $\a$.
For angular dependence $\xi^{(1)}$, $\alpha=\mu-1$, where
$\mu = [(j+1/2)^2 - q^2]^{1/2}$ and
$j=q+1/2,q+3/2,...$, while for angular dependences $\xi^{(2)}$ and $\eta$,
$\alpha=\mu$ with $j=q+1/2,q+3/2,...$ for $\xi^{(2)}$ and $j=q-1/2$ for
$\eta$. In all cases the multiplicity is $(2j+1)$.
In three dimensions, fermionic wave functions (or,
more precisely, wave sections \cite{WY,KYG}) $\Psi_F$ in (\ref{ferm})
are doublets. This doublet structure is carried by the angular
dependences, so both for bosons and fermions radial functions $\psi$ in
\eq{rad} are one-component objects. Therefore, unlike the
scattering problem in (3+1) dimensions \cite{KYG}, our calculation does
not require any special treatment of $j=q-1/2$ fermionic modes.

We can treat bosonic and fermionic cases simultaneously using \eq{rad}
if we adopt the following notation
\be
\alpha = [(j+1/2)^2 - q^2]^{1/2} - \kappa - 1/2 \; ,
 ~~~~~~~j=q+\kappa,q+\kappa+1,... \; .
\label{alpha}
\ee
Then, $\kappa=0$ corresponds to
bosons, $\kappa=1/2$ to fermions with angular dependence $\xi^{(1)}$,
and $\kappa=-1/2$ combines fermions with angular dependencies
$\xi^{(2)}$ and $\eta$. Results for the vacuum sector are obtained by
substituting $q=0$. Though such substitution leads to an unphysical
value $j=-1/2$ for $\kappa=-1/2$, eigenvalues corresponding to this
unphysical value do not participate in the fermionic trace in (\ref{S2})
because of the vanishing multiplicity factor $(2j+1)$.

By the change of variables
\be
\psi(r)=r^{\alpha}(1+r^2)^{-\alpha-1/2} \phi(x) \; ,
 ~~~~~x=(1+r^2)^{-1} \; ,
\label{change}
\ee
\eq{rad} is converted into a hypergeometric equation.
The resulting eigenvalues are
\be
\lambda_n = (n+\alpha + 1/2)(n +\alpha + 3/2) \; , ~~~n=0,1,... \; .
\label{eigen}
\ee
We still need an ultraviolet regularization for the traces in \eq{S2}.
A convenient one is provided, again in parallel with 't Hooft's
calculation \cite{tHooft}, by two Pauli-Villars regulators with masses
$M_i$ and metrics $e_i$ satisfying $\sum_i e_i = -1$,
{}~$\sum_i e_i M_i^2=0$,
{}~$i=1,2$. Regularized traces that we will need are
\be
\Tr\log {\cal M}(M_i;\kappa) =
\sum_{j=q+\kappa}^{\infty} (2j+1) \sum_{s=1}^{\infty} \sum_{i=0,1,2}
e_i \log[(s+\alpha)^2 + \mu_i^2] \; ,
\label{reg}
\ee
where we defined $e_0=1$, $M_0=0$, and $\mu_i^2 = M_i^2-1/4$,
{}~$i=0,1,2$.
The effective action in the fermionic case is obtained from
the half-sum of traces (\ref{reg}) with $\kappa=\pm 1/2$ which are
in fact related to each other in a simple way.

Because the effective action of monopole is dimensionless, it can
depend only on products of infrared and ultraviolet regulator
parameters, $M_i R$. In the system of units where $R=1$, we are then
interested in the dependence of the effective action on $M_i$ in the
limit when $M_i$ are large. Non-regulator terms in \eq{reg}
cannot produce such dependence. Applying the Euler-Maclaurin formula to
regulator terms in \eq{reg} we get
\[
\sum_{s=1}^{\Lambda} \log[(s+\alpha)^2 + M^2] =
\int_0^{\Lambda} ds \log[(s+\alpha)^2 + M^2]
\]
\be
+ \left. \left( \frac{1}{2} \log[(s+\alpha)^2 + M^2] +
       \frac{1}{6} \frac{s+\alpha}{(s+\alpha)^2+M^2}
\right) \right|^{\Lambda}_0 + O([\alpha^2+M^2]^{-3/2}) \; ,
\label{EM}
\ee
where $\Lambda \gg M$. The remainder in \eq{EM} gives a convergent
contribution of order $1/M$ when summed over $j$ with
the multiplicity factor $(2j+1)$ and hence may be neglected.
Terms divergent at $\Lambda\to\infty$ as well as those proportional to
$M_i^2$ get cancelled when all regulator and non-regulator contributions
are added together and we obtain
\[
\Tr\log {\cal M}(M_i;\kappa) = \sum_{j=q+\kappa}^{\infty} (2j+1) \times
\]
\be
\left[ \sum_{i=1,2} e_i
\left(
-(\alpha + 1/2)\log(\a^2+\mu_i^2) + 2 \mu_i \left( \frac{\pi}{2}-
\mbox{arctan}\frac{\a}{\mu_i} \right) -
\frac{1}{6}\frac{\a}{\a^2+\mu_i^2}
\right) + C(j) \right] \; ,
\label{summed}
\ee
where $C(j)$ denotes terms independent of $M_i$.

The dependence of \eq{summed} on $M_i$ can be found by the following
method. We decompose each sum over $j$ into two -- one running from
$q+\kappa$ to some value $J-1$ such that $q \ll J\ll M_i$, another running
from $J$ to a cutoff $\Lambda'\gg M$. As in the sum over $s$ before,
dependence on the cutoff will disappear when all regulator
and non-regulator terms are added together.
The number $J$ is integer or half-integer when $q+\kappa$
is integer or half-integer, respectively.
Now, in the region $q+\kappa \leq j \leq J-1$ we can
neglect $j$ compared to $M_i$ while in the region
$J\leq j \leq \Lambda'$ we can use the Euler-Maclaurin formula.
At $J\leq j \leq \Lambda'$ we can also use the expansion
\be
\alpha = \alpha_1 - q^2 (2j+1)^{-1} + O(j^{-3}) \; ,
 ~~~~~\alpha_1 = j - \kappa \; .
\label{exp}
\ee
For example,
\[
\sum_{j=q+\kappa}^{\Lambda'} (2j+1)(\a+1/2)\log(\a^2+M^2)
\]
\[
= \sum_{j=q+\kappa}^{J-1} (2j+1)(\a+1/2)\log M^2
+ \sum_{j=J}^{\Lambda'} (2j+1)(\a_1+1/2)\log(\a_1^2 + M^2)
\]
\be
- q^2 \sum_{j=J}^{\Lambda'} \left(
\log(\a_1^2 + M^2) + \frac{\a_1 (2\a_1 + 1)}{\a_1^2 + M^2} \right)
+ O(M^{-2}) + O(J^{-1}) \; .
\label{log}
\ee
It turns out that other terms in \eq{summed} do not produce contributions
in either boson or fermion determinant that distinguish between
monopole and vacuum sectors. Proceeding with \eq{log}, we finally obtain
the one-loop effective action of a monopole with monopole number $q$
in the presence of massless bosons and fermions up to terms
independent of $R$:
\[
S^{(1)}_{B,F}(R)= K_{B,F}(q) \log M^2 R^2  + O(R^0) \; ,
\]
\be
K_B(q) = \lim_{J\to\infty} \left[
\sum_{j=q}^{J-1} (2j+1) [(j+1/2)^2-q^2]^{1/2} -\frac{2}{3} J^3
+\frac{1}{6} J + q^2 J \right] \; ,
\label{resb}
\ee
\be
K_F(q) = - \lim_{J\to\infty} \left[
\sum_{j=q+1/2}^{J-1} (2j+1) [(j+1/2)^2-q^2]^{1/2}
- \frac{2}{3} J^3 + \frac{1}{6} J + q^2 J \right] -\frac{q}{2} \; .
\label{resf}
\ee
We assumed that both regulator masses are of the same order, $M_i\sim M$.
The limits in eqs.(\ref{resb})-(\ref{resf}) were done by computer.
The results for a few values of $q$ are presented in Table 1.
\begin{table}
\begin{tabular}{|r|r|r|} \hline
$q$ & $K_B(q)$ & $K_F(q)$ \\ \hline
1/2 & 0.0968 & 0.0151 \\
1   & 0.2266 & 0.1730 \\
3/2 & 0.3850 & 0.4358 \\
2   & 0.5682 & 0.7852 \\ \hline
\end{tabular}
\caption{Coefficients of $\log{R^2}$ in the one-loop effective action
of a monopole with monopole number $q$ in the presence of massless bosons
and fermions.}
\end{table}

When there are $N$ species of particles of given type, the
corresponding numbers from Table 1 should be multiplied by $N$.
For sufficiently large $N$, logarithms of $R$ coming from
boson or fermion determinants will overpower $3\log MR$ that comes
with the opposite sign from the volume factor.
Thus, when there is a large number of massless matter species, monopoles
are suppressed. For small $N$, we cannot draw any conclusions from the present
work because the one-loop calculation is not a reliable guide in this case.
Possibly, numerical simulations can help to solve the problem for small $N$.

Let us now state some applications of our results. They show that
non-confining abelian gauge interactions are possible in (2+1) dimensions
when gapless excitations are present, thus providing some support to the
idea of new gauge interactions in planar electronic systems. In a somewhat
different interpretation, our results provide a mechanism by which
interlayer tunneling of excitations with one unit of the ordinary electric
charge is suppressed while that of a doubly charged
object is allowed. Recent work \cite{tunnel} shows that a viable theory of
high-temperature superconductors can be constructed if such mechanism is
assumed to exist. Let us postulate
that a planar system describing one layer in a layered material
supports quasiparticles that carry magnetic flux with respect to the
new gauge field. There are two varieties of such quasiparticles
corresponding to positive and negative fluxes, respectively.
Assume further that these quasiparticles carry ordinary electric charge
("holons") which has the same sign both for positive and negative
fluxes. Events of interlayer tunneling of these quasiparticles are
described by monopoles and anti-monopoles in three dimensions.
If there are also excitations of another
sort ("spinons") that carry charge, rather than flux, with respect to the
new gauge field and are gapless, tunneling of a {\em single} flux can be
suppressed by their infrared effects as discussed above, while a {\em pair}
of positive and negative fluxes can tunnel freely.

I am grateful to S. Chakravarty for emphasizing to me the relevance
of monopoles to the problem of spin-charge separation and discussions
of the results and to V. Rubakov for discussions on fermion-monopole
interactions. It is a pleasure to acknowledge the hospitality of the
Aspen Center for Physics where this work was started. The author is
supported by the Julian Schwinger fellowship at UCLA.

\end{document}